%% file: main.tex
\documentclass[10pt]{article}
\usepackage[preprint]{tmlr}
\usepackage{tmlr}

\input{math_commands.tex}

\usepackage[utf8]{inputenc} %
\usepackage[T1]{fontenc}    %
\usepackage{hyperref}       %
\usepackage{url}            %
\usepackage{booktabs}       %
\usepackage{amsfonts}       %
\usepackage{nicefrac}       %
\usepackage{microtype}      %
\usepackage{xcolor}         %
\usepackage{graphicx}
\usepackage{natbib}
\usepackage{mdframed}
\usepackage{listings}    %
\usepackage{tcolorbox}
\usepackage{todonotes}
\usepackage{amsmath}
\usepackage{xspace}
\usepackage{wrapfig}
\usepackage{booktabs}
\usepackage{array}
\usepackage{listings}
\usepackage{tabularx}
\usepackage{caption}
\usepackage{rotating}

\title{Gradient-Based Program Repair:\\Fixing Bugs in Continuous Program Spaces}

\author{%
  André Silva \quad Gustav Thorén \quad Martin Monperrus \\
  KTH Royal Institute of Technology \\
  \texttt{\{andreans,gthor,monperrus\}@kth.se} \\
}

\author{\name André Silva \email andreans@kth.se \\
      \addr KTH Royal Institute of Technology
      \AND
      \name Gustav Thorén \email gthor@kth.se \\
      \addr KTH Royal Institute of Technology
      \AND
      \name Martin Monperrus \email monperrus@kth.se \\
      \addr KTH Royal Institute of Technology
}

\newcommand{\raspbugs}{RaspBugs\xspace}
\newcommand{\gbprfull}{Gradient-Based Program Repair (GBPR)\xspace}
\newcommand{\gbprlong}{Gradient-Based Program Repair\xspace}
\newcommand{\gbpr}{GBPR\xspace}
\newcommand{\totalRaspBugs}{1,466\xspace}

\begin{document}

\maketitle

\begin{abstract}
Automatic program repair seeks to generate correct code from buggy programs, with most approaches searching the correct program in a discrete, symbolic space of source code tokens.
This symbolic search is fundamentally limited by its inability to directly reason about program behavior.
We introduce \gbprfull, a new approach that recasts program repair as continuous optimization in a differentiable numerical program space.
Our core insight is to compile symbolic programs into differentiable numerical representations, enabling search in the numerical program space directly guided by program behavior.
To evaluate \gbpr, we present \raspbugs, a new benchmark of \totalRaspBugs buggy symbolic RASP programs and their respective numerical representations.
Our experiments demonstrate that \gbpr{} can effectively repair buggy symbolic programs by gradient-based optimization in the numerical program space, with convincing repair trajectories.
To our knowledge, we are the first to state program repair as continuous optimization in a numerical program space.
Our work demonstrates the feasibility of this direction for program repair research, bridging continuous optimization and program behavior.
\end{abstract}

\section{Introduction}

Program repair, or automatic bug fixing, promises to generate corrective patches for faulty code \citep{monperrus2018automatic}.
Recent years have seen dramatic improvements in the quality and complexity of patches thanks to learning based program repair, with the most complex bugs being repaired by frontier LLMs \citep{yang2024swe}.
Progress on benchmarks like SWE-Bench \citep{jimenez2024swe} and RepairBench \citep{silva2024repairbench} has demonstrated that real-world bugs can be fixed automatically.

The fundamental limitation of asking a language model to generate a patch is that it does so by reasoning about token distributions, and not by reasoning about the expected behavior. 
In other words, optimizing for next token prediction captures very little of the difference between buggy behavior and expected correct behavior from the specification.
For the same reason, it is hard to repair completely new programs as language models fail to generalize to unseen problems \citep{chollet2024arc}.

In this paper, we completely reframe learning based program repair.
We propose to embed the specification and the incorrect behavior to be repaired as a first-class concept in a loss function that is used to directly optimize the program.

We describe \gbprfull{}, a novel approach that is founded on expressing programs as numerical representations, such as embeddings or neural network weights. With this numerical program representation associated with a loss that captures the expected behavior, \gbpr repairs the bug by searching in the numerical program space.
This core originality of \gbpr is that it considers program behavior, the expected correct one and the buggy one, as a first-class concept in the learning pipeline, directly expressed in the loss function to be optimized. 
To sum up, we propose to 1) transform symbolic programs into numerical programs, 2) design loss functions that capture correct behavior, and 3) optimize numerical programs with gradient descent in the program space until a repair is found.

\begin{figure}[t]
    \centering
    \includegraphics[width=\linewidth]{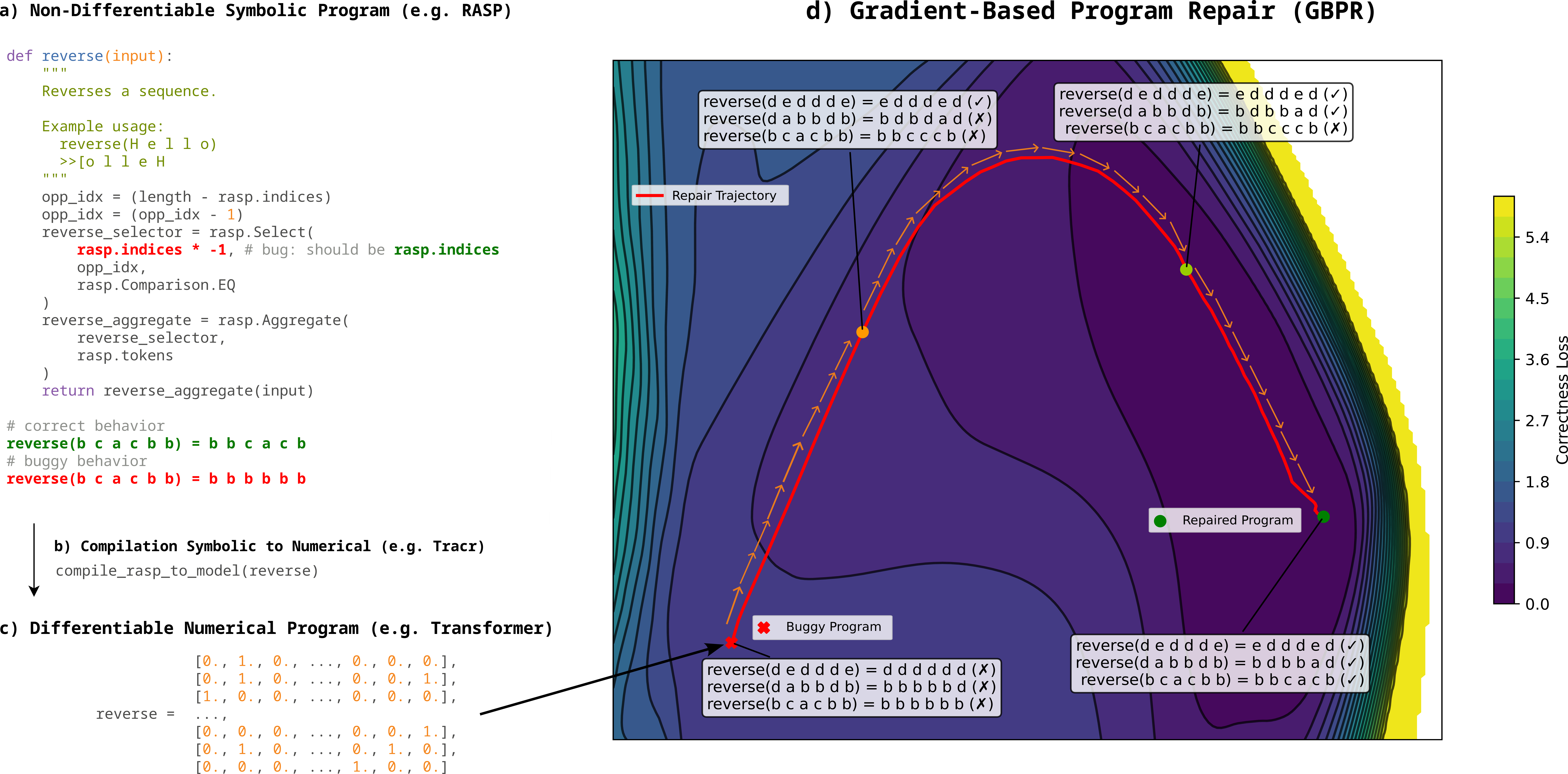}
    \caption{
    The key insight of \gbprlong{} is that program search can be done in a numerical space by employing gradient-based optimization.
    a) Symbolic program computing the \texttt{reverse} function, written in RASP, and the difference between the expected and buggy behavior;
    b) Compilation of the symbolic program into a numerical program, encoded as a Transformer;
    c) Numerical program, equivalent to the symbolic program;
    d) \gbpr optimizes the numerical program via the correctness loss, starting from the buggy program.
    The program is iteratively optimized, moving towards correct behavior. As the correctness loss decreases, the program correctness increases, with some incorrect behavior now corrected.
    At the end of the optimization, the repaired program correctly implements the \texttt{reverse} function.
    As opposed to LLM-based bug fixing, \gbpr directly reasons about the expected behavior as a first-class optimizable concept.
    }
    \label{fig:raspair}
\end{figure}

To rigorously evaluate our approach, we introduce \raspbugs, a new benchmark of buggy symbolic programs and their corresponding numerical representation. Those programs are written in RASP \citep{weiss2021thinking}, a class of sequence-processing programs that can be analytically represented as Transformer models.
By systematically applying mutations to base RASP programs, we create a diverse collection of meaningful bugs that can be analyzed both at the symbolic and numerical levels.
\raspbugs contains \totalRaspBugs{} bugs over 6 programs, and provides the first-ever controlled environment for researching program repair as continuous optimization.

Our results demonstrate that gradient-based program repair is feasible.
First, we observe proper convergence of the optimization problem, with the correctness of the considered programs improving.
Second, we are able to repair the majority of buggy programs for 5 out of the 6 considered base programs.
Third, the analysis of the repair trajectories and the correctness landscape confirms that incorrect behavior on buggy input points is gradually fixed.

To summarize, our main contributions are:
\begin{itemize}
    \item \gbprlong{}, a novel approach for program repair that performs gradient-based program search, driven by a loss function that captures correct program behavior.
    \item \raspbugs{}, a curated benchmark for evaluating research on continuous program repair, with \totalRaspBugs pairs of buggy RASP programs, available as either symbolic or numerical programs.
    \item An empirical evaluation demonstrating the feasibility and effectiveness of \gbpr{} for repairing RASP programs as continuous optimization in the numerical program space.
\end{itemize}

\section{Background}
\label{sec:background}

\textbf{Program Repair.}
Program repair \citep{monperrus2018automatic} automatically finds a correct program from a buggy one, changing incorrect behavior to correct behavior according to a specification (e.g., an input-output test suite), typically via search or mutation over the program space.
Most program repair research considers repairing imperative programs, in particular Python or Java.

\textbf{Symbolic Program Space.}
In the context of traditional program repair, programs are symbolic artifacts, represented using discrete structures like source code token sequences, abstract syntax trees (ASTs), or control-flow graphs (CFGs).
Program repair on symbolic programs relies on symbolic methods operating directly on these structures according to rules based on language syntax and semantics (e.g., program transformations, static analysis, symbolic execution).
Large Language Models (LLMs) are used for program repair \citep{vasic2018neural, yasunaga2021break, yang2024swe} by considering code as a sequence of textual tokens.

\textbf{Numerical Program Space.}
A numerical program is a program 1) whose behavior is encoded as continuous, real-valued parameters and 2) can be executed.
These can be either neural networks or vectors in latent spaces with execution semantics, such as in \cite{bonnet2024searching}.
Unlike traditional symbolic programs, which are constrained by discrete structures, the behavior of numerical programs can be adjusted smoothly via optimization techniques like gradient descent.

\textbf{RASP Language.}\label{sec:prelim-rasp-tracr}
RASP (Restricted Access Sequence Processing) is a domain-specific programming language for sequence processing~\citep{weiss2021thinking}. Its key caharacteristic is that the language primitives align with the transformer-encoder model.
RASP programs operate on token and position streams using selectors (\(n\times n\) relations that hold when a predicate over per-position values is true), per-position aggregation (reductions along selectors), and elementwise maps.
We adopt the bounded setting of RASP which assumes fixed vocabulary and a maximum sequence length.

\textbf{Tracr Compiler.}
Tracr \citep{lindner2023tracr} is an analytical, deterministic compiler from RASP to encoder-style transformer models.
It programmatically constructs attention from RASP selectors (select \(\to\) query/key scores high exactly where the predicate holds; values route information) and implements aggregation and maps via attentiona and MLP steps respectively.
For a fixed vocabulary and sequence length, the compiled model is strictly equivalent to the RASP program (\(D(\mathbf{x}) = P(\mathbf{x})\) on the supported domain).

\section{Gradient-Based Program Repair}
\label{sec:dpr}

All previous research has done program repair as a search in a symbolic space.
Our core insight is that one can do program repair by searching programs in a numerical space instead.
In that numerical space, the program semantics are encoded into a numerical representation.
\gbprfull{} leverages gradient descent to search the numerical program space, minimizing a loss that directly measures deviations from correct behavior.
The program zeroing the loss is considered the repaired program.

\subsection{Compilation of Symbolic Programs to Differentiable Numerical Programs}

The first step of numerical repair is to translate the initial symbolic program into a numerical representation where a correctness gradient can be computed with respect to its parameters.
Let \( P_f \) be a symbolic program (e.g., source code text in Python) that implements the target function \( f: \mathcal{X} \rightarrow \mathcal{Y} \), mapping inputs from space \( \mathcal{X} \) to outputs in space \( \mathcal{Y} \).

\textbf{Compilation.} We require a compiler function, denoted \( \mathcal{C} \), that transforms \( P_f \) into a numerical representation \( D_{f, \theta} \).
This representation \( D_{f, \theta} \) is parameterized by a set of numerical parameters \( \theta \), such that executing the numerical representation on an input \( \mathbf{x} \in \mathcal{X} \) yields the program's output.
Crucially, the compiler \( \mathcal{C} \) must ensure that the numerical parameters \( \theta \) completely encode the semantics of the original program \( P_f \). In other words, \( \theta \) `is' the numerical program.

\textbf{Numerical Execution.} The execution of \( D_{f, \theta} \) must match the input-output behavior of \( P_f \) on the supported domain, guaranteeing that program semantics agree in both the symbolic and numerical spaces.
\[
D_{f, \theta} \equiv P_f \implies D_{f, \theta}(\mathbf{x}) = P_f(\mathbf{x}) \quad \forall \mathbf{x} \in \mathcal{X}.
\]

\textbf{Differentiation.}
We require \( D_{f, \theta} \) to be differentiable over \(\mathcal{X}\) with respect to \(\theta\).
This means we need to compute the gradient of a loss function, in order to change the parameters \( \theta \) to improve the correctness of the output for \(\mathbf{x}\).
If the gradient captures correctness, this means that gradient descent is actually optimizing the program towards more correct behavior, which is the fundamental goal of program repair (\autoref{sec:background}). 

\textbf{Alternatives for \( D_{f, \theta} \).}
In \autoref{sec:discussion}, we will discuss a few appropriate representations for \( D_{f, \theta} \).
At this point, we focus on neural networks as our numerical representation.
The neural network input, resp. output, is the program input, resp. output.
This is a natural choice as 1) neural networks are inherently differentiable via backpropagation, 2) their parameters form the continuous space \( \theta \) we seek to optimize, and 3) they are executable via forward passes.

\subsection{\gbprfull{}}

Let us assume a buggy symbolic program \( P_b \) implementing an incorrect function \( b \).
The ideal correct function is called  \( f \), and is defined by a specification that describes the behavior of the ideal program.
In this paper, we assume specifications in the form of input-output examples: \( \{(\mathbf{x}_i, \mathbf{y}_i)\}_{i=1}^n \), where each input \( \mathbf{x}_i \) is mapped to its correct output \( \mathbf{y}_i \) by the ideal function \( f \).

Symbolic repair means directly changing \( P_b \) with e.g., symbolic repair templates or repair operators that manipulate symbols.
\gbpr{} means repairing the numerical representation \( D_{f, \theta} \) instead.
For this, we first compile  \( P_b \) using \( \mathcal{C} \) to obtain its differentiable representation \( D_{b, \theta_b} \).
Both the initial parameters \( \theta_b \) and the structure of the numerical program are given by the compiler.
The goal of \gbpr{} is to adjust these parameters \( \theta_b \) to find a new set of parameters \( \theta^* \) such that the behavior of \( D_{\theta^*}(\mathbf{x}) \) matches the specification.

\[
D_{\theta^*}(\mathbf{x}) = f(\mathbf{x}) \quad \forall \mathbf{x} \in \mathcal{X}.
\]

\textbf{Correctness Loss.}
Next, we need a loss function \( \mathcal{L} \) that measures how far the current program behavior deviates from the specification.
The total loss is an aggregation of a local loss function \( \ell \) computed over a subset of the specification:
\[
\mathcal{L}(\theta, \{(\mathbf{x}_i, \mathbf{y}_i)\}_{i=1}^n) = \sum_{i=1}^n \ell\left(D_{\theta}(\mathbf{x}_i), \mathbf{y}_i\right).
\]

Consider the space of all possible parameter values \( \theta \) for our differentiable numerical program \( D_{\theta} \).
Each point in this space corresponds to a slightly different program behavior.
The loss function \( \mathcal{L} \) creates a landscape over this space, where lower values indicate behavior closer to the correct program \( P_f \).

The repair process is then a classical optimization problem: finding the parameters \( \theta^* \) that minimize the correctness loss:
\[
\theta^* = \arg\min_\theta \mathcal{L}(\theta, \{(\mathbf{x}_i, \mathbf{y}_i)\}).
\]

\textbf{Repair as Gradient Descent.}
Gradient descent acts like rolling a ball down this landscape. The initial parameters \( \theta_b \) place the ball somewhere corresponding to the buggy program's behavior.
The gradient \( \nabla_{\theta} \mathcal{L} \) points uphill towards higher loss (more incorrect behavior).
By moving in the opposite direction (\(-\nabla_{\theta} \mathcal{L}\)), we iteratively adjust the parameters \( \theta \), effectively improving the program's behavior step-by-step towards the desired correct functionality defined by the input-output specification.
Starting from the initial parameters \( \theta^{(0)} = \theta_b \) obtained from compiling the buggy program, we iteratively update the parameters in the direction opposite to the gradient of the loss:
\[
\theta^{(t+1)} = \theta^{(t)} - \eta \nabla_{\theta} \mathcal{L}(\theta^{(t)}),
\]
where \( \eta \) is the learning rate.

The main difference between symbolic repair and repair as gradient descent is that, because the representation  \( D_{\theta} \) is continuous and differentiable, small improvements are possible and efficiently guided by the gradient. This sharply contrasts with symbolic repair, which entirely consists of discrete jumps in the program space.

\subsection{Repair Acceptance Criterion}

Minimizing loss on training examples is insufficient for successful repair, as optimization might overfit, leading to a program \( D_{\theta^*}(\mathbf{x}) \) that performs well on training data but fails to generalize to unseen inputs and thus hasn't captured \( f \)'s true semantics.

Therefore, we need a repair acceptance criterion based on the performance of the optimized program \( D_{\theta^*} \) on a separate, held-out set of test examples \( \{(\mathbf{x}_j', \mathbf{y}_j')\} \) that were not used during the gradient descent optimization.
We consider the program repaired if its correctness on this held-out set exceeds \( 1 - \epsilon \) of the held-out test cases, for some small \( \epsilon \geq 0 \), ensuring that:

\[
D_{\theta^*}(x) \approx f(x) \quad \forall x \in \mathcal{X}.
\]

This ensures that the repair generalizes beyond the training data and the program likely corresponds to the intended function.

\subsection{Summary of Key Novel Concepts}

\textbf{Differentiable Numerical Programs.} Symbolic programs translated to continuous, differentiable forms (e.g., neural networks) with parameters (\(\theta\)) encoding semantics; a novel concept in the program repair literature.

\textbf{Numerical Repair Search Space.} Viewing the repair search space \(\theta\) as a continuous landscape where program behavior can be smoothly varied, as opposed to the irregular, discrete symbolic search space.

\textbf{Correctness Loss.} A differentiable function \(\mathcal{L}\) quantifying the difference between the current program's behavior \(D_{\theta}(\mathbf{x})\) and the expected behavior \(\mathbf{y}\). We cast classical optimization loss into a behavioral semantics conceptual framework.

\textbf{Correctness Gradient.} \(\nabla_{\theta} \mathcal{L}\), indicating the direction in numerical program space towards correct behavior.

\textbf{\gbprlong{}.} Iteratively adjusting program parameters \(\theta\) via gradient descent on the correctness loss (\(\theta^{(t+1)} = \theta^{(t)} - \eta \nabla_{\theta} \mathcal{L}\)), optimizing towards functional correctness. This is the first framing of program repair as continuous optimization, in contrast to traditional discrete symbolic search.

\section{\raspbugs: A Benchmark of Buggy Transformer Programs}
\label{sec:raspbugs}

\begin{wrapfigure}{R}{0.5\textwidth}
    \centering
    \includegraphics[width=0.5\textwidth]{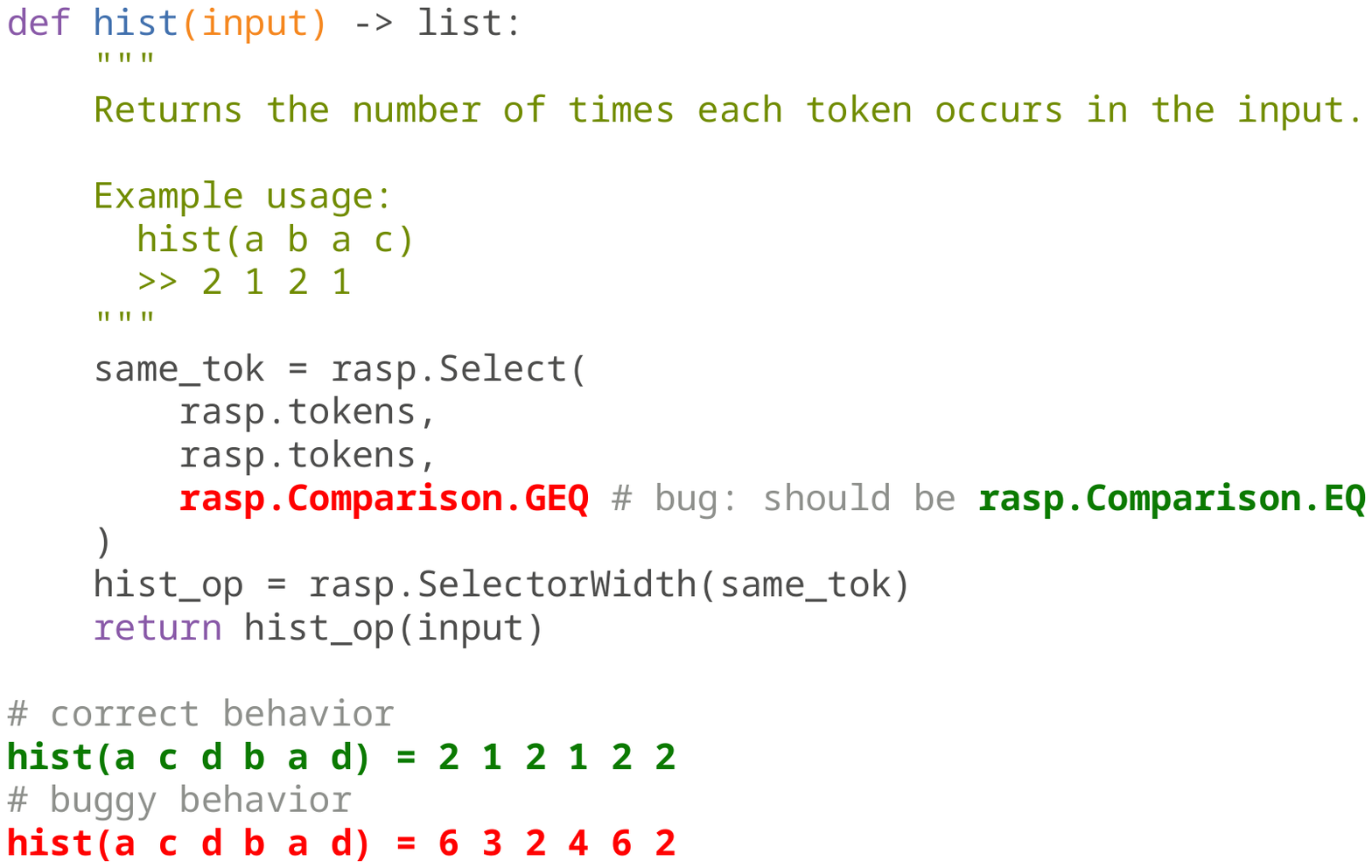}
    \caption{
        Example of a buggy RASP program in \raspbugs{}, synthesized from the reference \texttt{hist} program using mutation.
        The reference program selects only equal tokens, while the mutated program selects tokens greater than or equal to, resulting in buggy program behavior.
    }
    \label{fig:mutation_example}
\end{wrapfigure}

To evaluate \gbpr{}, we need buggy symbolic programs and their equivalent differentiable numerical counterparts.
We thus build \raspbugs{}, a novel benchmark of buggy transformer programs.
We choose to consider RASP programs \citep{weiss2021thinking}, which have the property to be representable symbolically or as Transformer models (see \autoref{sec:background})

\textbf{Programs.}
We rely on previous work by \cite{weiss2021thinking} and six of their reference RASP programs.
These programs perform various sequence processing operations, including sorting, reversing, histogram computation, frequency-based sorting, and validating Dyck language expressions.

\textbf{Input-Output Specifications.}
For each RASP program, we generate an input-output specification by randomly sampling from the input space and computing the corresponding outputs using the ground-truth symbolic implementation.
Each program specification is composed of 50,000 I/O pairs.
The lengths of the input samples are randomly sampled between 2 and 10.
Each specification is split into train ($80\%$), validation ($10\%$), and test ($10\%$) sets.

\textbf{Mutating Transformer Programs.}
We create \raspbugs{} by applying a suite of mutation operators to the original RASP programs.
The mutations are meant to introduce semantic changes to the program.
We consider generic mutation operators that act on programming language operators such as arithmetic operations and comparison operations.
We also design and implement nine RASP-specific mutation operators that target constructs of the RASP language.
In total, we utilize 15 mutation operators.
These mutation operators are employed individually or combined with others to generate higher-order mutants - mutated programs with several changed locations.
We set the limit of mutations per order per program to 200.

\textbf{Properties of Mutated Programs.}
Buggy programs must:
1) be symbolically buggy (at least one input-output pair is incorrect),
2) compile to Transformer models via Tracr \citep{lindner2023tracr},
3) be executable numerically (forward pass), and
4) be numerically buggy (incorrect on the same input-output pairs).
Validation outcomes include: \texttt{FAILED\_MUTATION} (symbolic interpretation errors), \texttt{UNCOMPILABLE} (Tracr compilation failure), \texttt{CORRECT\_MODEL} (semantically equivalent mutations), and \texttt{BUGGY\_MODEL} (programs for repair, considered hereafter).

\textbf{Descriptive Statistics.}
\raspbugs{} is composed of \totalRaspBugs{} buggy RASP programs, seeded from six reference programs and 15 mutation operators, their corresponding input-output specifications (split into train, validation, and test sets), and their numerical representations as Transformer models.
The buggy programs are broken to a different extent, as demonstrated by their different test set accuracies: min = 0.00\% (completely broken), median = 2.00\%, average = 36.69\%, max = 98.00\% (a corner-case bug).
The numerical representations range from 2k (\texttt{hist} program) to 1M (\texttt{dyck2} program) parameters.
Full details about \raspbugs{} can be found in \autoref{app:raspbugs}.

\section{Experiments}
\label{sec:experiments}

\begin{figure}[t]
  \centering
  \includegraphics[width=\textwidth]{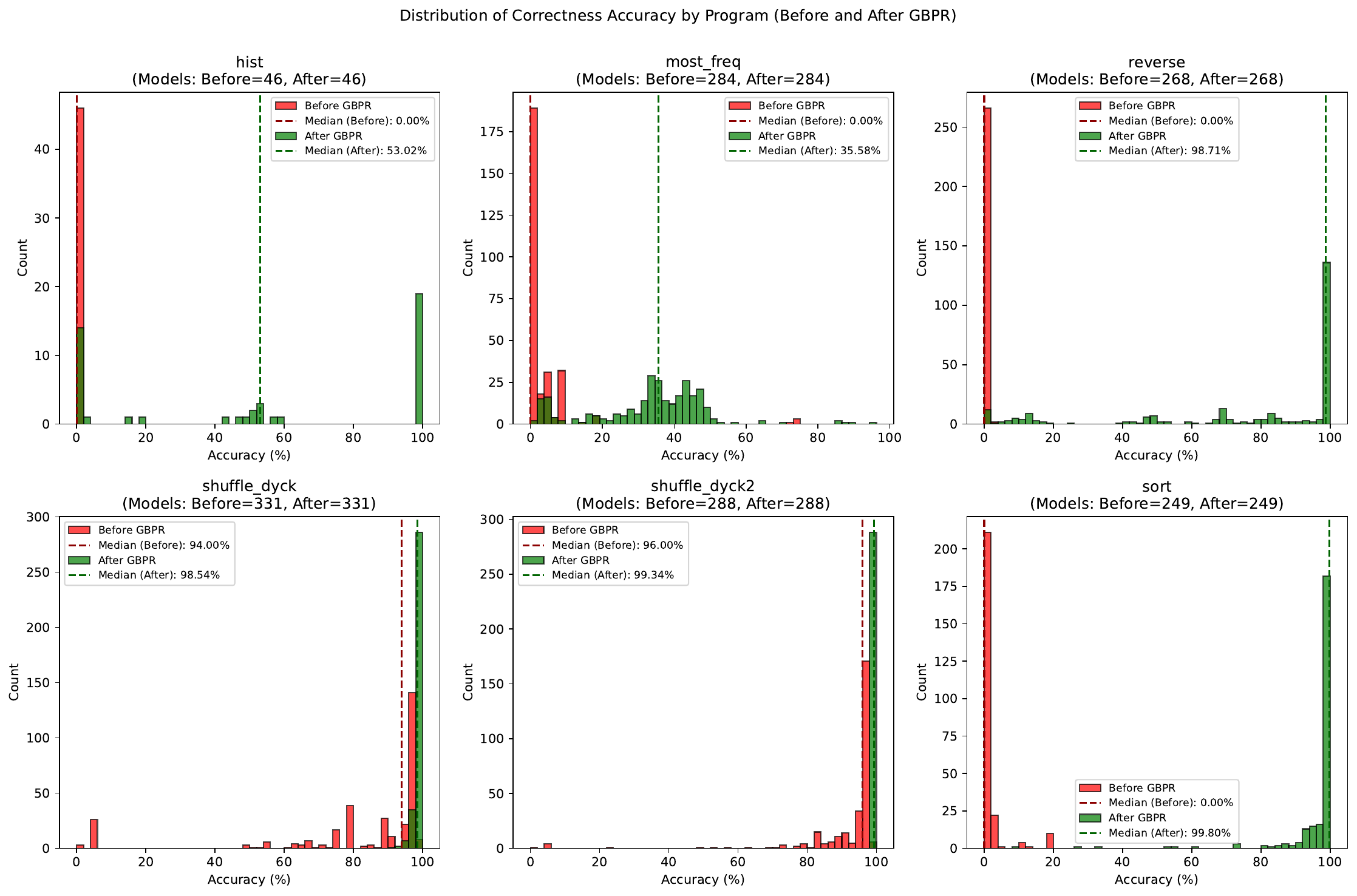}
  \caption{Accuracy distribution before (red) and after (green) \gbprlong{} for each program in \raspbugs{}. The majority of buggy variants for five programs can be repaired with \gbpr{} (as demonstrated by the rightmost green bars).}
  \label{fig:acc-dist-per-program}
\end{figure}

\subsection{Training and Evaluation.}
Each buggy transformer program is fine-tuned via supervised learning on its train split (\autoref{sec:raspbugs}), minimizing cross-entropy correctness loss between predicted and ground-truth output sequences.
We use batch size 256, learning rate \(1 \times 10^{-4}\), and train up to 10k epochs with early stopping (validation loss improvement \(< 1 \times 10^{-4}\) for 10 epochs).
Repaired programs are evaluated on the test set via greedy decoding (temperature 0), reporting accuracy as exact output match percentage.
Experiments used multi-instance NVIDIA A100 GPUs (1/7th A100 compute, 10GB VRAM, 2 CPUs, 32GB RAM per instance/run).

\subsection{Repairing Transformer Program.}
To evaluate the effectiveness of \gbprlong{}, we apply it to the entire \raspbugs{} benchmark.
Our goal is to determine whether gradient-based optimization can reliably repair a wide variety of buggy transformer programs.

\autoref{fig:acc-dist-per-program} shows the correctness accuracy over the test sets for the buggy programs before and after \gbprlong.
Here, correctness accuracy is defined as the percentage of test samples for which the model's output exactly matches the ground-truth output.
For example, the top-left figure shows the correctness accuracy distribution over 46 buggy hist programs from \raspbugs{}.
The red distribution shows that most buggy programs are completely broken with a correctness accuracy of close to 0\%.
The green distribution represents the correctness accuracy after repair.
We see that a large number of hist programs have higher correctness after \gbprlong (green distribution shifted to the right), with the majority achieving near perfect correctness (right-most bar).

Before repair (red bars), for four of the six program types, the majority of buggy numerical programs start with near-zero correctness accuracy (red bars clustered at 0\%).
This indicates that the mutations introduce substantial semantic errors, resulting in programs that almost never produce correct outputs.

After repair (green bars), the accuracy distribution shifts dramatically to the right for five out of six program types.
In all these cases, the majority of repaired programs achieve near-perfect correctness, demonstrating that \gbprlong{} can repair incorrect behavior even for severe bugs (i.e., those with initial accuracies near 0\% as detailed in \autoref{sec:raspbugs}).

For the \texttt{most-freq} program, while correctness clearly improves, most programs do not achieve perfect accuracy after repair.
This suggests inherent difficulties for gradient-based methods with certain programs, possibly due to complex loss landscapes or significant architectural changes (a point further discussed in \autoref{sec:discussion}).

Overall, our experiments over \totalRaspBugs buggy transformer programs clearly demonstrate the viability of \gbprlong. It is effective to use gradient optimization and an input-output specification to repair a broken symbolic program in the RASP/Tracr setting, establishing a proof-of-concept for continuous program repair.

\begin{tcolorbox}[colback=gray!5,colframe=black,title=\textbf{Takeaway}]
\gbpr{} successfully repairs buggy programs in the numerical program space. Experiments show it restores correctness for the majority of bugs across 5 out of 6 base RASP programs, achieving near-perfect repair even for initially completely broken programs.
\end{tcolorbox}

\begin{figure}[t]
    \centering
    \includegraphics[width=0.329\linewidth]{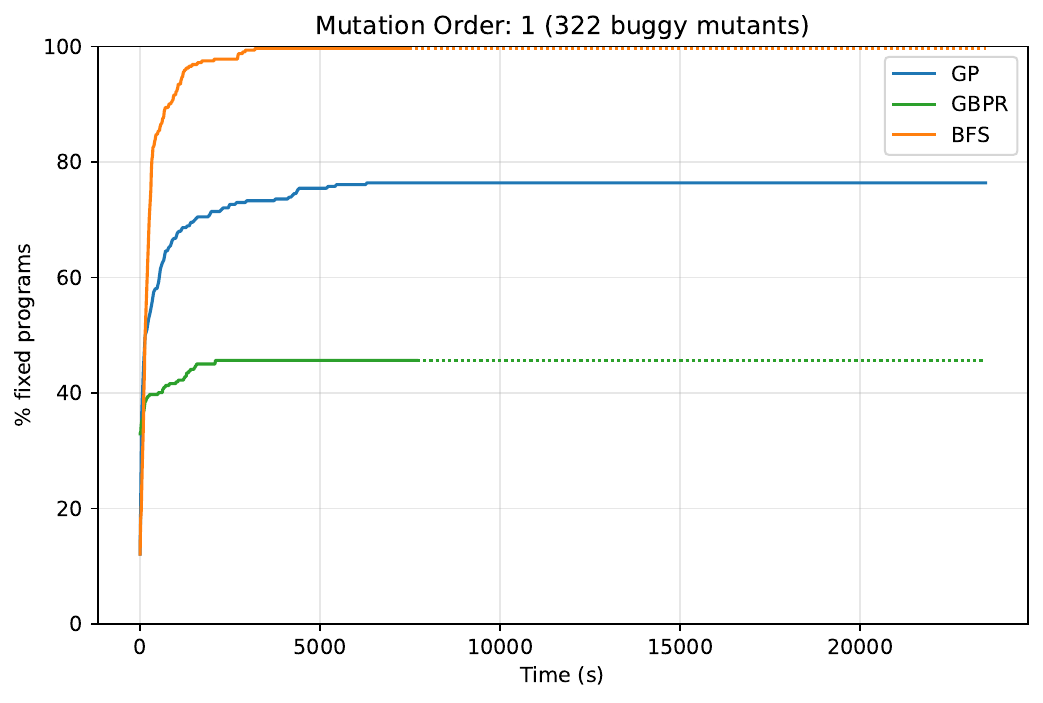}
    \hfill
    \includegraphics[width=0.329\linewidth]{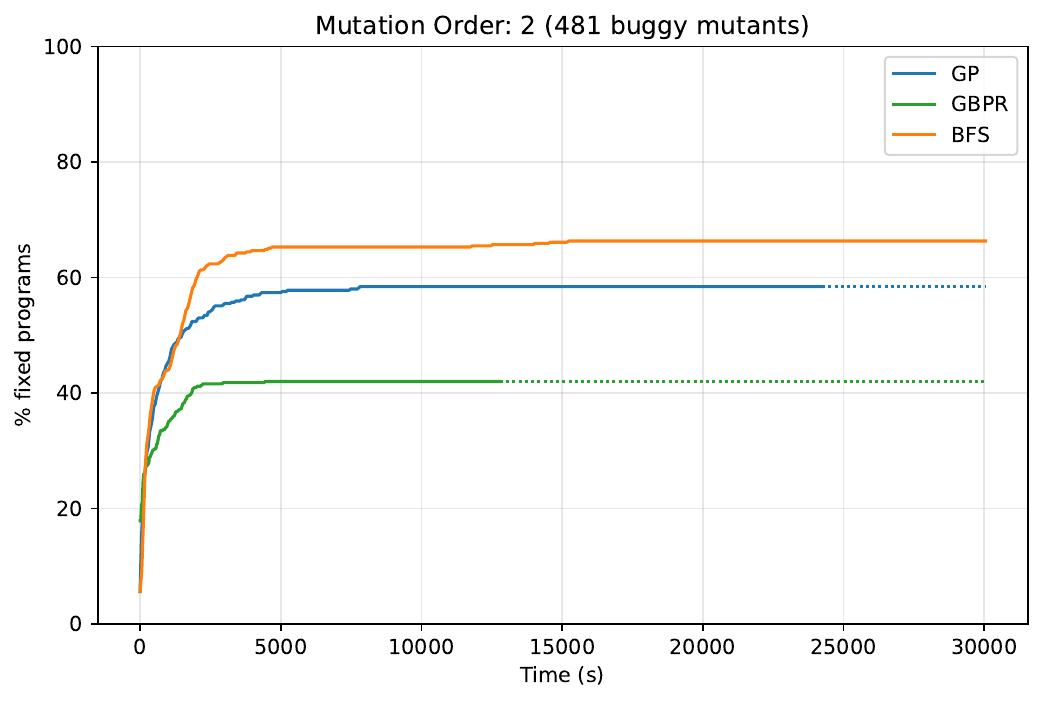}
    \hfill
    \includegraphics[width=0.329\linewidth]{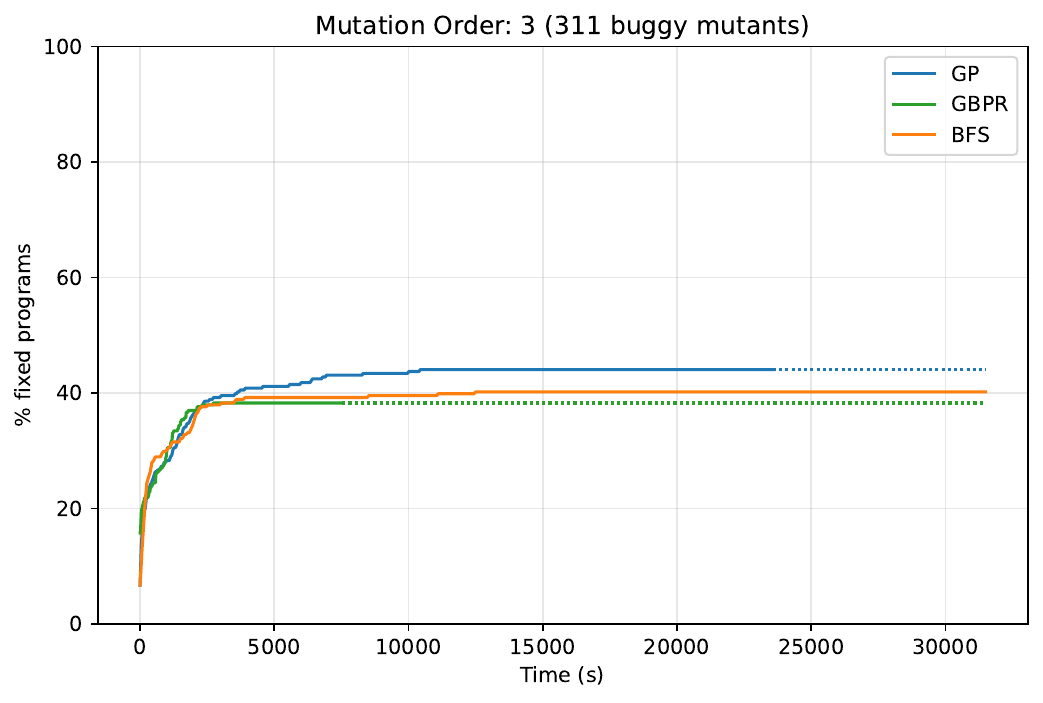}

    \vspace{0.5em}

    \hspace{\linewidth}
    \includegraphics[width=0.329\linewidth]{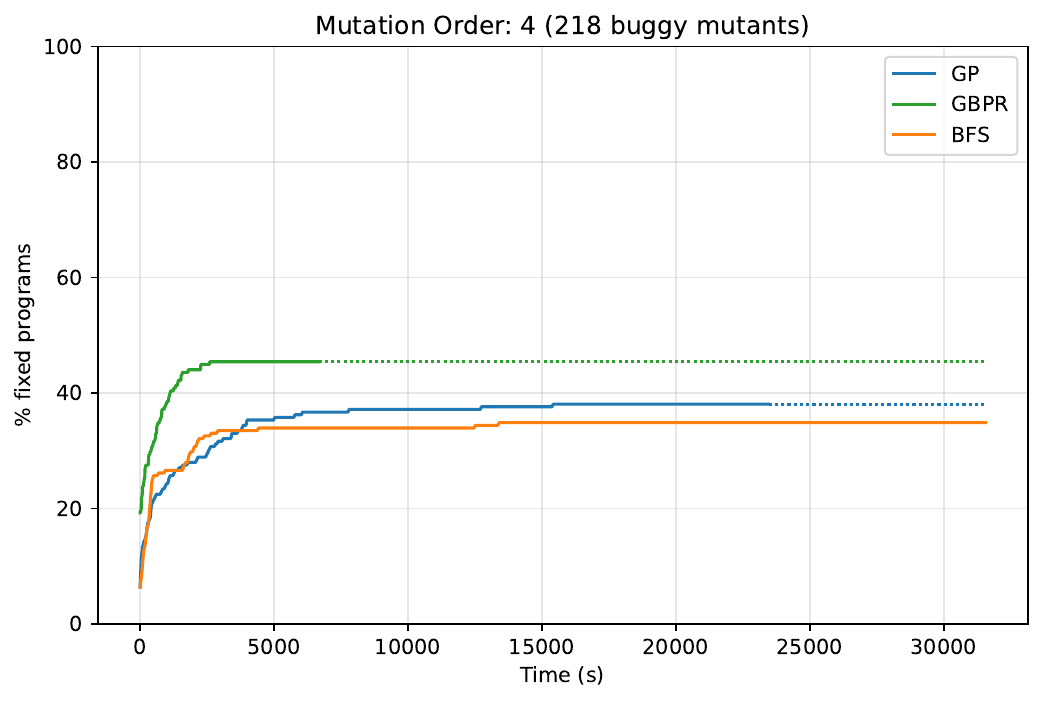}
    \hspace{0.02\linewidth}
    \includegraphics[width=0.329\linewidth]{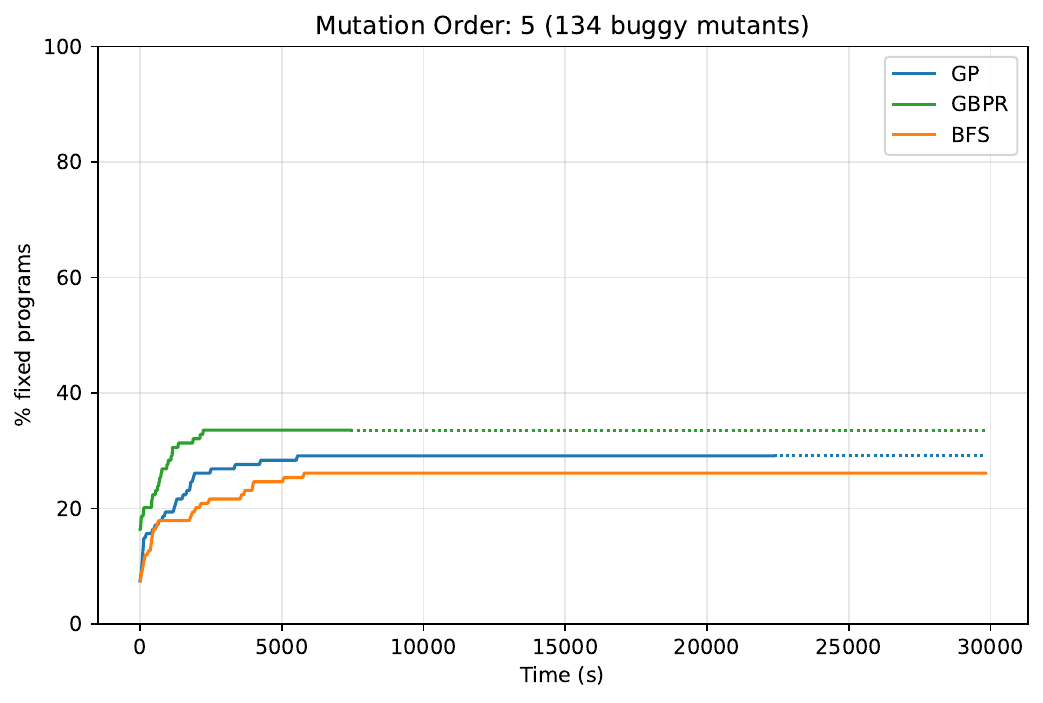}
    \caption{Repair success rates (test accuracy $\geq 99\%$) over evaluation steps, stratified by mutation order.
    The x-axis counts gradient steps for \gbpr{} and programs evaluated for the symbolic baselines; a wall-clock time comparison is provided in \autoref{fig:time-comparison}.
    GP is run with a budget of 1,000 programs per bug.
    Each panel shows pass-rate trajectories for \gbpr{}, GP, and BFS (higher is better).
    For simple bugs (orders 1--2), symbolic methods achieve higher final performance since the number of potential repairs to explore is small.
    For complex bugs (orders 4--5), \gbpr{} surpasses both symbolic baselines, showing how gradient-based optimization scales better with bug complexity.}
    \label{fig:baseline-comparison}
\end{figure}
\subsection{Comparison with Symbolic Baselines.}
To contextualize the effectiveness of \gbprlong{}, we compare it against two baselines that operate in the symbolic program space, as opposed to \gbpr{} which operates in the numerical program space.

The first baseline employs a Genetic Programming (GP) approach with an evolutionary strategy with population size $\mu=16$ and offspring generation $\lambda=16$ per generation, operating directly on RASP program source code represented as Abstract Syntax Trees (ASTs).
Starting from the buggy program, the initial population is seeded with the original bug plus random single-step mutations.
Each generation proceeds as follows: (1) parent selection via tournament selection with $k=3$ (i.e., we select 3 programs at random and keep the one with the highest fitness), (2) mutation of selected parents to generate offspring according the mutation operators, and (3) replacement where the top $\mu$ individuals by fitness are retained from the combined pool of parents and offspring.
The fitness function is defined as accuracy on the training dataset, computed by executing each candidate program on all input-output pairs and measuring sequence-level exact matches (ignoring the BOS token).
The search terminates either when achieving 100\% accuracy or after evaluating 1,000 programs (approximately 30 complete generations). We chose this budget after verifying that GP's progress curve plateaus well before the 1,000-program mark on the majority of bugs, indicating diminishing returns from additional evaluations.
The second baseline implements exhaustive breadth-first search (BFS) that systematically explores all possible mutations paths from the buggy program.

Critically, both baselines use the same mutation operators employed to generate the buggy programs in \raspbugs{} (\autoref{sec:raspbugs}). This represents an optimistic baseline: in realistic repair scenarios, one would not have prior knowledge of the exact fault operators that introduced the bugs, whereas our baselines benefit from this oracle information.

\begin{figure}[t]
    \centering
    \includegraphics[width=0.329\linewidth]{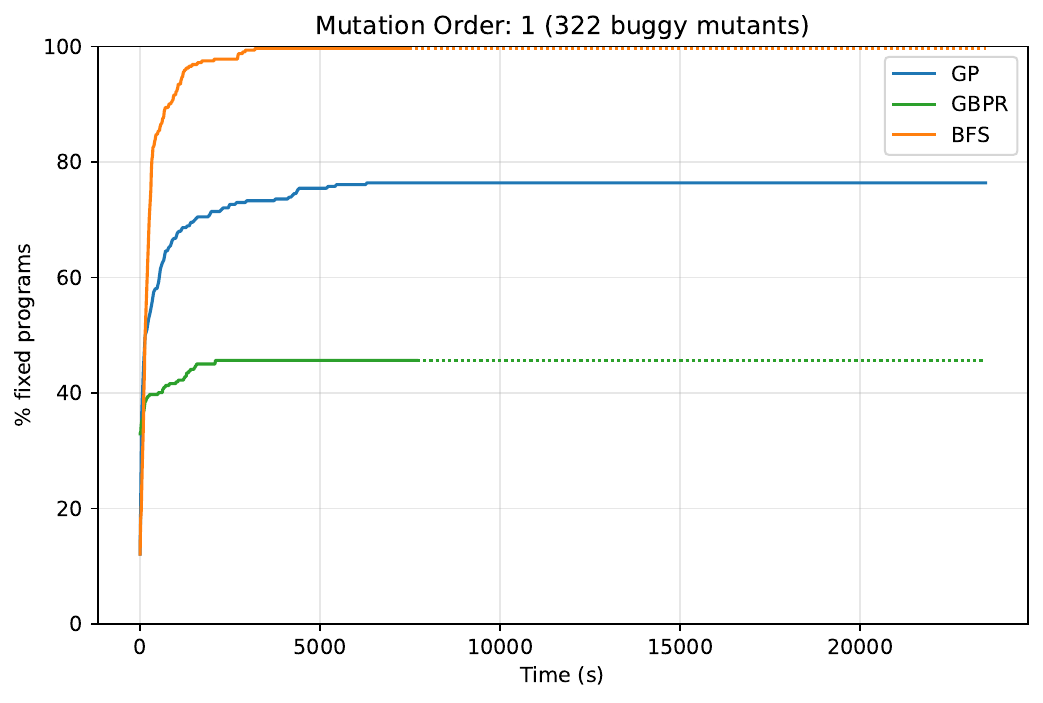}
    \hfill
    \includegraphics[width=0.329\linewidth]{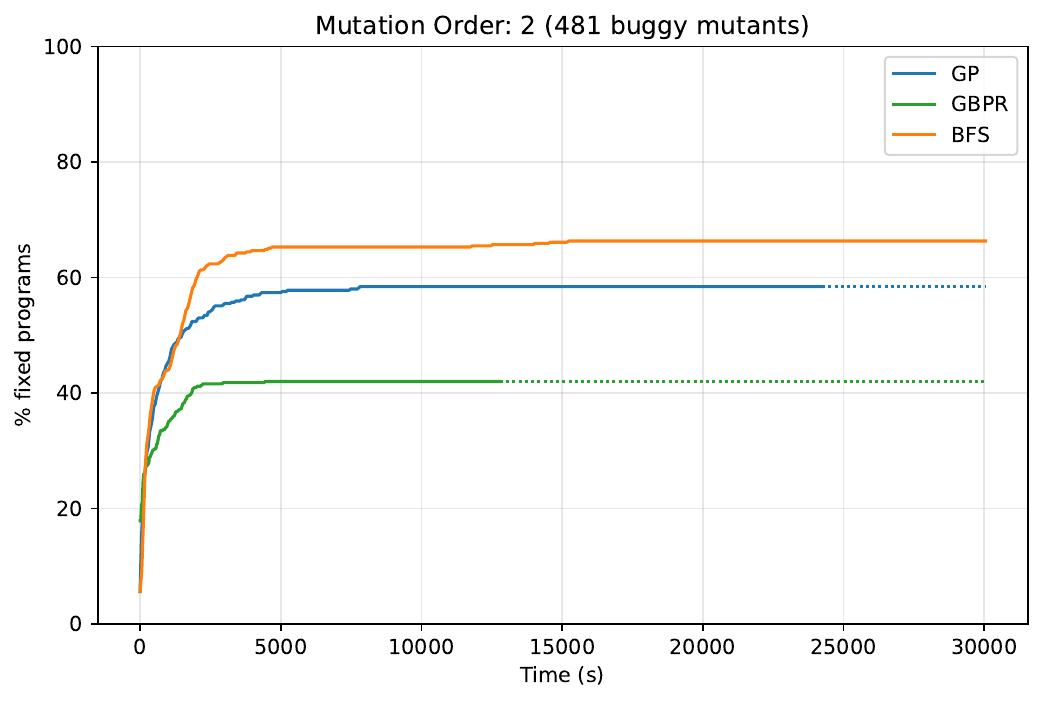}
    \hfill
    \includegraphics[width=0.329\linewidth]{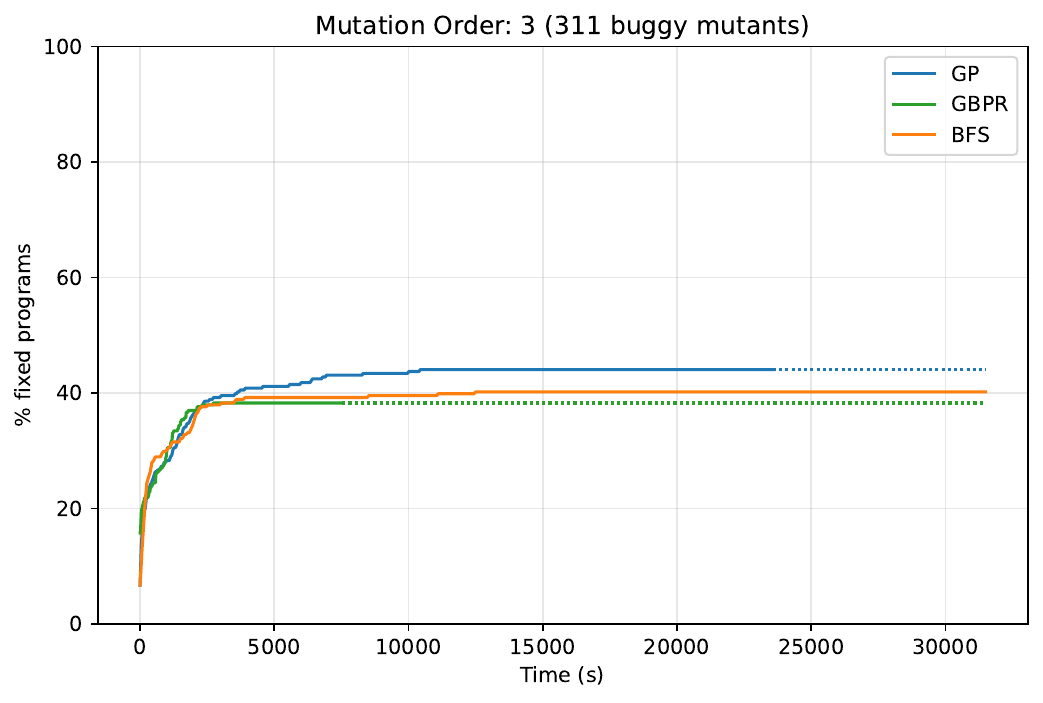}

    \vspace{0.5em}

    \hspace{\linewidth}
    \includegraphics[width=0.329\linewidth]{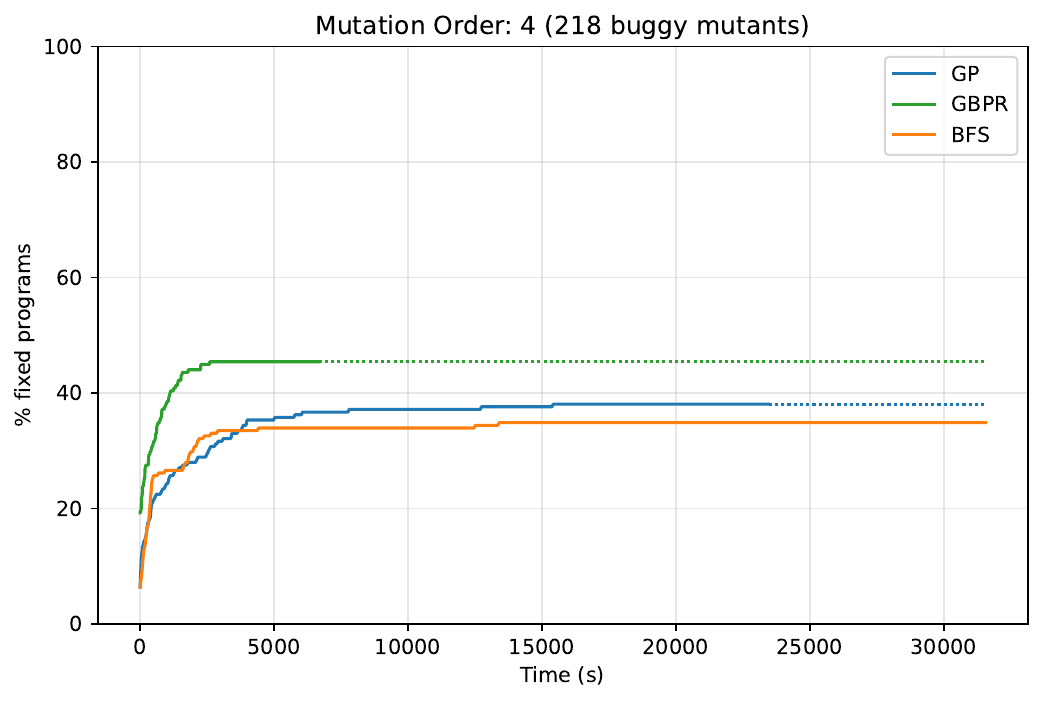}
    \hspace{0.02\linewidth}
    \includegraphics[width=0.329\linewidth]{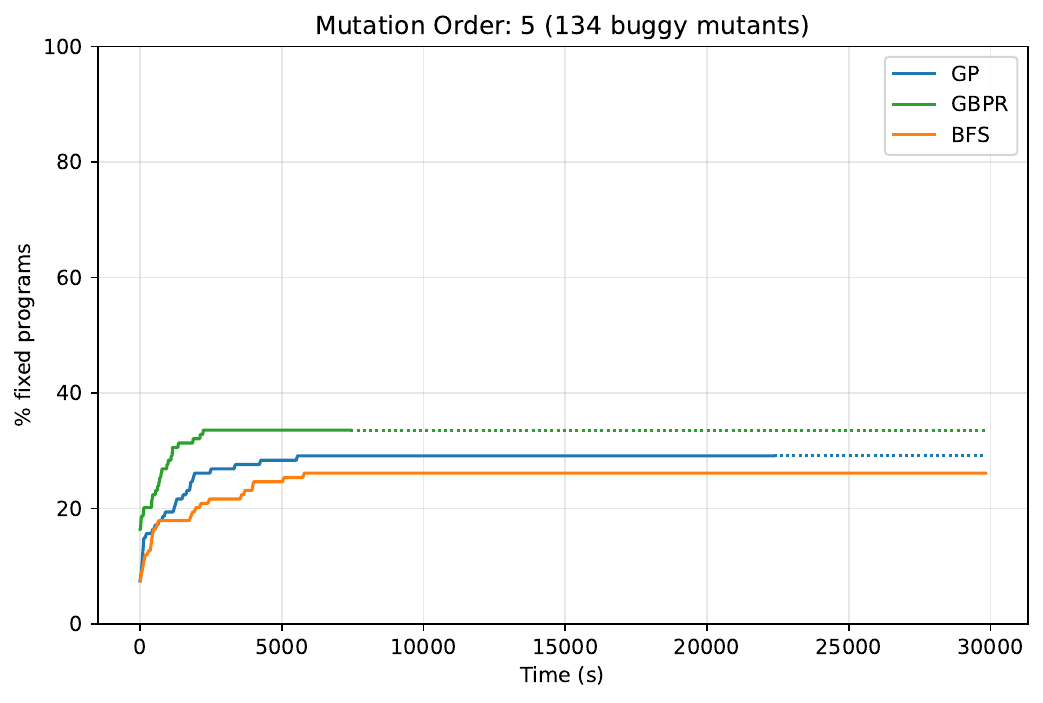}
    \caption{Repair success rates (test accuracy $\geq 99\%$) over wall-clock time, stratified by mutation order.
    Each panel shows pass-rate trajectories for \gbpr{}, GP, and BFS (higher is better).
    For simple bugs (orders 1--2), symbolic methods retain their advantage.
    For complex bugs (orders 4--5), \gbpr{} leads both GP and BFS at every time cutoff; \gbpr{} reaches its final performance around 1 hour, while GP and BFS continue improving before plateauing later.}
    \label{fig:time-comparison}
\end{figure}

\autoref{fig:baseline-comparison} shows repair pass rates (test accuracy $\geq 99\%$) over evaluation steps, stratified by mutation order (one panel per order: single mutations in the top left, five simultaneous mutations in the bottom right).
Each panel plots pass rate trajectories for all three methods, revealing both convergence dynamics and final performance.
\autoref{fig:time-comparison} provides the corresponding wall-clock time comparison, which is the more principled measure of compute efficiency.

\textbf{Asymptotic behavior.} First, we see that all three methods exhibit asymptotic behavior after some time.
\gbpr{} asymptotically reaches a performance level after approximately 75 training epochs, while the two symbolic baselines continue improving through their respective evaluation budgets but at slower rates.
This reflects fundamentally different search dynamics. Gradient-based optimization is able to quickly make use of the training signals via gradient, combining the information of the behavior for multiple I/O samples at once. On the contrary, symbolic methods do not have this powerful joint convergence over I/O samples, but instead are able to steadily explore the search space.

\textbf{Performance by bug complexity.} Examining convergence patterns across mutation orders reveals how different paradigms scale with bug complexity.
For simple bugs (mutation orders 1-2), symbolic methods achieve substantially higher pass rates at convergence (top left and middle figures).
BFS fixes all single-mutation bugs (order 1, top left) as it simply explores all single-step mutations from the buggy program, which is to be expected and also acts as a sanity check for our pipeline.
As the number of mutations increases, the number of potential repairs grows exponentially, meaning that symbolic search becomes ever harder. 
Indeed, at mutation order $\geq 3$, \gbpr{} starts
to be competitive compared to symbolic search, with faster optimization but still lower final performance (top right).
For complex bugs (orders 4-5), \gbpr{} surpasses both symbolic baselines at their respective convergence points.
This clearly validates the intuition that \gbpr{} enables a joint optimization over multiple faulty locations, where a single move along the gradient in the numerical space is equivalent to multiple mutations in the symbolic space.

This pattern reflects fundamental differences in how those two search paradigms (symbolic versus numerical) scale with bug complexity.
Symbolic methods are able to systematically explore the search space of small modifications, excelling at bugs requiring one ot two local changes.  However, they face combinatorial challenges as the number of potential repairs grows exponentially with the number of locations to fix.
Continuous optimization navigates high-dimensional parameter spaces via gradients, trading performance on local, simple bugs for the ability to handle complex, multilocation bugs via moves in the numerical space.

\begin{tcolorbox}[colback=gray!5,colframe=black,title=\textbf{Takeaway}]
While symbolic search methods (GP, BFS) excel at simple bugs, \gbpr{} outperforms them on complex, multi-location bugs. Gradient-based optimization scales better with bug complexity by enabling joint updates across the entire program structure in the numerical program space.
\end{tcolorbox}

\textbf{Wall-Clock Time Comparison.}
The step-based view above does not account for the different per-step costs of the three methods.
GBPR training uses GPU acceleration (1/7th A100 per run), while GP and BFS are CPU-only, so wall-clock time is a more principled but also more hardware-dependent comparison.
\autoref{fig:time-comparison} shows the resulting wall-clock time curves across all six base programs.

The advantage of \gbpr{} is most pronounced for complex bugs.
For mutation orders 4 and 5, \gbpr{} leads both GP and BFS at every time cutoff.
\gbpr{} reaches its final performance around 1 hour (after which all jobs have finished training via early stopping): 45.4\% for order-4 bugs and 33.6\% for order-5 bugs.
GP and BFS continue to improve beyond 1 hour but plateau later --- for order-4 bugs, GP reaches 38.1\% and BFS 34.9\% by $\sim$6 hours; for order-5 bugs, both plateau around 2 hours (GP: 29.1\%, BFS: 26.1\%).
At their respective convergence points, \gbpr{} maintains a 7--8 percentage-point advantage over the next-best method on order-4 bugs, and a 4--5 percentage-point advantage on order-5 bugs.

\begin{tcolorbox}[colback=gray!5,colframe=black,title=\textbf{Takeaway}]
Under a wall-clock time budget, \gbpr{} strictly outperforms both GP and BFS on complex, multi-location bugs (orders 4--5), where symbolic search faces a combinatorial explosion. \gbpr{} converges around 1 hour; symbolic methods plateau later but at lower final performance.
\end{tcolorbox}

\subsection{Effect of Specification Size.}
\label{sec:spec-size}

A natural question is how sensitive \gbpr{} is to the number of input-output examples used during repair.
We conducted an ablation over $N \in \{100, 1{,}000, 5{,}000, 10{,}000, 25{,}000, 40{,}000\}$ specification examples, using the same gradient-step budget (500k steps) and early-stopping mechanism as in the main experiments.
Three regimes emerge clearly: below $N < 5{,}000$, symbolic methods dominate at all mutation orders; at $N \approx 10{,}000$, \gbpr{} becomes competitive on complex bugs (orders 4--5), matching or surpassing GP and BFS; and at $N \approx 25{,}000$--$40{,}000$, \gbpr{} clearly surpasses both symbolic baselines on complex bugs.
For simple bugs (orders 1--2), symbolic methods retain their advantage throughout the tested range, as the repair search space is small and gradient descent offers no advantage there.
Full results, including a per-order breakdown and figure, are provided in \autoref{app:spec-size}.

\begin{tcolorbox}[colback=gray!5,colframe=black,title=\textbf{Takeaway}]
\gbpr{} is competitive on complex bugs already at $N \approx 10{,}000$ examples, well below the full specification size. For simple bugs, symbolic methods remain more data-efficient. The crossover point is cleanly connected to the mutation order.
\end{tcolorbox}

\begin{figure}[t]
    \vspace{-1cm}
    \centering
    \begin{minipage}[c]{0.49\linewidth}
        \centering
        \includegraphics[width=\linewidth]{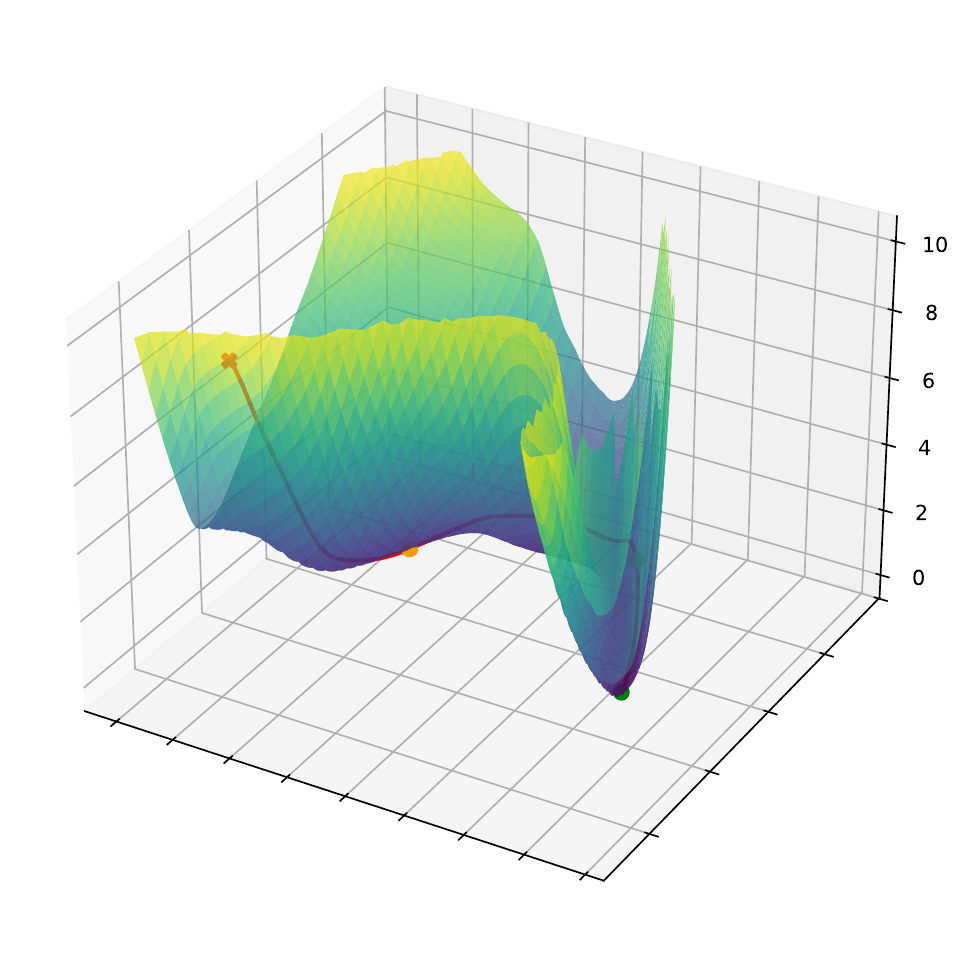}
    \end{minipage}
    \hfill
    \begin{minipage}[c]{0.49\linewidth}
        \centering
        \includegraphics[width=\linewidth]{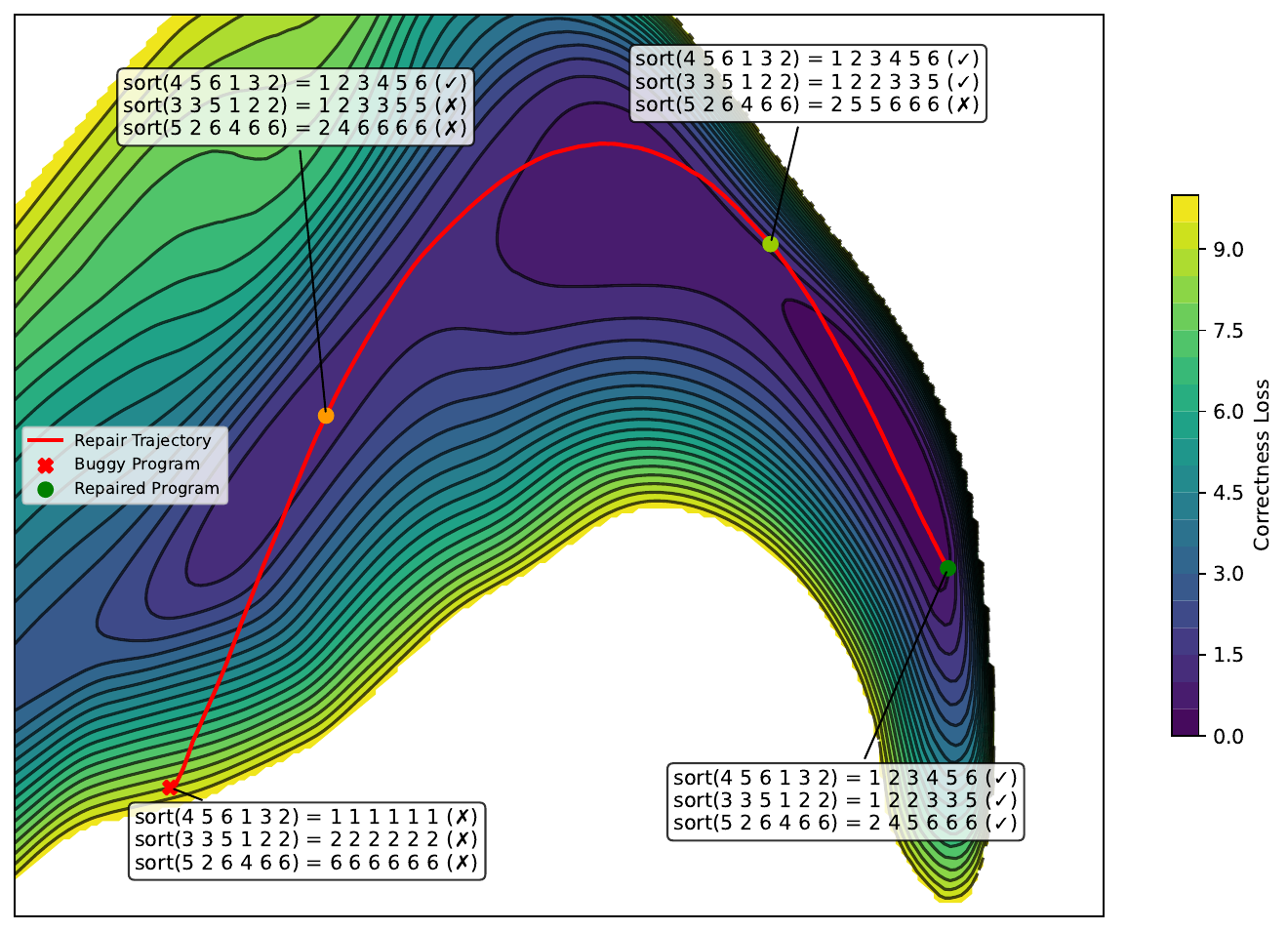}
    \end{minipage}
    \caption{
    Repair trajectory for a buggy \texttt{sort} program, in the numerical program space.
    The red cross marks the initial buggy program, and the trajectory shows the path taken by gradient descent towards a repaired program.
    \gbprlong{} iteratively updates the numerical representation of the program using the gradient defined by the correctness loss landscape, until the program behavior is repaired ($L \approx 0$).
    Left: Surface plot of the correctness loss landscape along the two principal components of the numerical program space. Right: Contour plot of the same landscape, the input-output behavior changing along the trajectory.
    }
    \label{fig:sort-example}
\end{figure}

\subsection{Repair Trajectories through the Correctness Landscape.}
To provide further insight into how \gbprlong{} operates, we visualize repair trajectories from buggy programs to repaired ones across the numerical program space.

\autoref{fig:sort-example} shows the repair trajectory for a buggy \texttt{sort} program, and the surrounding correctness landscape. In this landscape, higher loss means more buggy behavior.
The left panel presents the surface plot while the right panel shows a contour plot for the same trajectory, augmented with input-output behavior sampled from the trajectory.
The red cross indicates the starting point of the search, i.e., the buggy program encoded as a numerical program, which has high loss and low correctness.
As \gbpr{} proceeds, the program is iteratively updated, following the steepest descent in the loss landscape.
The trajectory ultimately converges to a minimum, where the program is successfully repaired and near-perfect correctness on the test set.

From an execution perspective, at the beginning, the buggy \texttt{sort} program (red cross) is not capable of sorting any of the three input examples.
For example, an incorrect output lists the same element multiple times.
During repair, the program gradually improves.
At the second highlighted point, the program already correctly sorts the first example.
However, at this point, the repair is only partial -- the remaining two examples are not correctly sorted -- which is reflected by the relatively high loss.
At the third highlighted point, the program correctly sorts two of the examples, with the loss now closer to 0.
As the loss landscape is explored, the program eventually converges to a minimum where the loss is minimized and the accuracy maximized.
This means that the program is successfully repaired and behaves according to the provided specification.

This visualization highlights the core novelty of our approach: by representing programs as differentiable objects, we exploit the topology of the loss landscape to guide the repair process via gradient descent towards correct behavior. This is in sharp opposition to relying on discrete, combinatorial search for repairing symbolic programs.

In summary, our experimental results demonstrate that \gbprlong{} is feasible, it can reliably repair a wide range of buggy transformer programs, often achieving near-perfect correctness.
The approach is robust across different RASP programs and various bugs seeded via different mutations.
Repair trajectories clearly demonstrate the repair dynamics happening in the numerical program space.

\begin{tcolorbox}[colback=gray!5,colframe=black,title=\textbf{Takeaway}]
Repair trajectory visualizations confirm that \gbpr{} meaningfully navigates the numerical program space. By following the gradient of the correctness loss, the optimization process steadily moves the program from buggy to correct behavior, with intermediate steps being progressively closer to specification.
\end{tcolorbox}

\section{Discussion}
\label{sec:discussion}

\textbf{Specification Types in the Loss Function.}
A key concept of gradient-based program repair is that it expresses the specification in the loss. Hence, the gradient directly captures the program's incorrect behavior.
In our experiments, we have used the cross-entropy loss over an input-output specification, appropriate for the considered RASP programs.
Ultimately, \gbpr opens the door to incorporating other rich behavioral information into the loss function, such as formal specifications or invariants.

\textbf{Differentiable Numerical Program Representations}.
Programs can be represented numerically in several ways.
In our experiments, we focus on neural networks, specifically Transformer models, as the numerical representation, compiled from symbolic RASP programs \citep{weiss2021thinking,lindner2023tracr,shaw2024alta}.
Other approaches include embedding programs as points in a continuous latent space, so-called latent programs, which also support efficient search and repair via continuous optimization methods \citep{bonnet2024searching}.
Execution of these numerical programs is performed by an auxiliary interpreter model.
Future work will focus on the design of advanced differentiable numerical representations that are ever more expressive.

\textbf{Decompilation to the Symbolic Space.}
A key future direction is decompiling repaired numerical programs into human-readable symbolic code.
Symbolic representation is both 1) more interpretable and amenable to human review and 2) appropriate for traditional verification techniques with guarantees.
However, this decompilation process is nontrivial: mapping the optimized parameters of a numerical representation back to structured, high-level code is an open research challenge, akin to decompilation.
Recent work has begun to address this problem in the context of Transformer models by discretizing the model \citep{friedman2023learning} or by training a meta-model to decompile weights into symbolic programs \citep{thurnherr2024neural,langosco2024towards}.
However, robust and general decompilation from neural programs to symbolic programs remains an unsolved research problem.

\textbf{Limitations.}
Our evaluation is limited to \raspbugs{}, our benchmark of RASP programs; broader experimentation with other RASP programs (e.g., with \cite{thurnherr2024tracrbench}) and languages is left to future work.
\gbpr{} can only optimize parameters within the initial model architecture obtained after Tracr compilation (see scope of Tracr in \autoref{sec:prelim-rasp-tracr}). If repairing a bug requires changing the structure itself (e.g., adding a new attention head), our prototype could not repair the bug. Future work is needed on symbolic-to-numerical compilation to maximize the expressivity of the numerical program space.
Another limitation is the reliance on input-output datasets to define the correctness loss. Our specification-size ablation (\autoref{app:spec-size}) shows that below $N < 5{,}000$ examples, symbolic methods are more effective at all mutation orders; \gbpr{} only becomes competitive on complex bugs (orders 4--5) at $N \approx 10{,}000$.
Our approach thus requires a test input generator that is good enough to explore the input space at this scale.

\textbf{Compiler Assumption.}
The RASP/Tracr setting is deliberately designed as a controlled environment in which programs compile analytically and exactly to transformer weights, removing the gap between the symbolic and numerical representations.
This is a significant simplification relative to general-purpose languages like Python, where no such analytical compiler currently exists.
The key question for future work is, therefore, how to construct differentiable numerical representations for broader program families, either via learned surrogates, smooth interpretation \citep{chaudhuri2010smooth}, or latent program spaces \citep{bonnet2024searching, macfarlane2026gradientbased}.
In this paper, we make no claim that \gbpr{} directly generalises to arbitrary languages; rather, we establish its feasibility as a proof-of-concept in a setting where the compilation assumption holds by construction, and articulate it as a research direction.

\section{Related Work}

\textbf{Latent Programs.}
Latent programs are represented in a latent space, a compressed feature space preserving meaningful data features and placing similar points adjacently.
\cite{Neelakantan2015NeuralPI} train a Neural Programmer to recursively select operations and data sources via latent representations at each execution step.
\cite{hong2021latent} find that generating discrete latent codes representing high-level operations improves program synthesis accuracy when compared with token-level generation.
\cite{bonnet2024searching} learn a latent program space for ARC-AGI programs, and use gradient-based search to find correct programs.
\cite{liskowski2020program} train an autoencoder to embed programs in a latent space, mapping them back with an evolutionary algorithm.
None of these works do program repair. Beyond our focus on RASP in this paper, \gbprlong is conceptually applicable to other latent program representations such as the ones from this related work.
Lastly, \cite{balog2020neural} introduce a neural synthesizer with a differentiable fixer that iteratively revises symbolic programs using latent program representation intermediates; unlike \gbpr{}, their method relies on the discretization of each edit step rather than optimizing the program fully in a continuous space.

\textbf{Learning Program Execution.}
Related work explores how neural networks can understand \citep{reed2015neural, shin2018improving, yan2020neural, chen2021latent} or benefit from \citep{ye2022neural, liu2023code} program execution.
For example, \cite{zaremba2014learning} learn LSTM networks to execute short Python programs.
\cite{ni2024next} teach models to inspect and reason about code execution by bootstrapping a synthetic training set of execution-aware reasoning traces.
In contrast to these works, which simulate execution with a black-box network, \gbpr{} expresses program behavior as a first-class concept within a numerical framework.

\textbf{Symbolic vs Numerical Program Spaces.}
The mapping between symbolic and numerical program spaces is a key component in gradient-based program repair, and a topic of earlier research.  
Neural surrogates \citep{esmaeilzadeh2012neural, renda2021programming} are neural networks designed to approximate complex programs and are typically trained on a subset of the input-output space.
\cite{weber2024learning} learn to compile source programs directly into neural surrogates, bypassing the need for generating input-output examples.
Smooth interpretation \citep{chaudhuri2010smooth} takes a different route: it smooths the semantics of a symbolic program to expose continuous parameters amenable to gradient-based optimization.
Other works focus on mapping from numerical to symbolic program spaces.
\cite{cranmer2020discovering} propose symbolic regression for extracting explicit symbolic models from learned models by relying on inductive biases matching to nature of the problem domain.
\cite{thurnherr2024neural, langosco2024towards, upadhyay2025model} study the decompilation of Transformer models into symbolic programs by learning meta-decompiler models.
Given such as mapping, the core novelty of \gbpr{} is to explore the numerical program space for finding patches, using gradient descent.

\textbf{Learning-based Program Repair.}
Several works have proposed using machine learning to repair programs \citep{long2016automatic, vasic2018neural, chen2019sequencer}.
In particular, LLMs are used to repair programs both in single-turn \citep{xia2023automated, jiang2023impact} and agentic \citep{yang2024swe, wang2024openhands} setups.
Our work is different in that we focus on repairing programs in a numerical space, using gradient-based optimization in search of the correct program, rather than searching exclusively in the token space.

\section{Conclusion}
\label{sec:conclusion}

We introduced \gbprfull{}, casting program repair as continuous optimization.
By compiling symbolic programs to differentiable numerical representations and applying gradient descent to a correctness loss, \gbpr{} effectively repairs buggy programs.
This work demonstrates the feasibility of expressing program execution semantics in a continuous, optimizable space, and opens new avenues for tackling fundamental programming problems via numerical optimization.

\bibliographystyle{plainnat}
\bibliography{references}

\newpage
\appendix
\input{appendix.tex}

\end{document}

%% file: math_commands.tex
\usepackage{amsmath,amsfonts,bm}

\def\eqref#1{equation~\ref{#1}}

\def\1{\bm{1}}

\DeclareMathAlphabet{\mathsfit}{\encodingdefault}{\sfdefault}{m}{sl}
\SetMathAlphabet{\mathsfit}{bold}{\encodingdefault}{\sfdefault}{bx}{n}

%% file: appendix.tex
\section{\raspbugs{}: Benchmark Details}
\label{app:raspbugs}

This appendix provides a detailed overview of the \raspbugs{} benchmark, designed to facilitate research in gradient-based program repair of Transformer programs.
\raspbugs{} consists of a collection of buggy RASP (Restricted Access Sequence Processing) programs, their corresponding correct versions, input-output specifications, and their compiled Transformer model representations.

The benchmark is built upon six base RASP programs, originally presented by \cite{weiss2021thinking}.
These programs cover a set of sequence-processing tasks, ranging from simple operations like sorting and reversing sequences to more complex tasks such as histogram computation and Dyck language validation.
Table~\ref{tab:rasp-programs} lists these base programs, along with a brief description and illustrative input-output examples for each.

\begin{table}[h]
    \centering
    \resizebox{\linewidth}{!}{%
    \begin{tabular}{
        l
        >{\raggedright\arraybackslash}p{0.32\linewidth}
        >{\raggedright\arraybackslash}p{0.18\linewidth}
        >{\raggedright\arraybackslash}p{0.18\linewidth}
    }
    \toprule
    \textbf{Program} & \textbf{Description} & \textbf{Example Input} & \textbf{Example Output} \\
    \midrule
    \texttt{sort} & Returns the input tokens sorted in ascending order. & \texttt{[1,5,3,4,3]} & \texttt{[1,3,3,4,5]} \\
    \addlinespace
    \texttt{reverse} & Returns the input sequence in the reverse order. & \texttt{[a,b,b,e,d]} & \texttt{[d,e,b,b,a]} \\
    \addlinespace
    \texttt{hist} & Returns the histogram count for each token in the input. & \texttt{[a,b,b,e,d]} & \texttt{[1,2,2,1,1]} \\
    \addlinespace
    \texttt{most-freq} & Returns the input sorted according to the token frequency in descending order. Only the first occurrence of each token in the output list is considered. & \texttt{[2,3,4,3,2,5]} & \texttt{[2,3,3,2,4,5]} \\
    \addlinespace
    \texttt{dyck-1} & Returns a sequence of ones if the sequence is balanced in regards to open and closed parenthesis, otherwise returns zeros. & \texttt{[(,),(,(,)]} & \texttt{[0,0,0,0,0]} \\
    \addlinespace
    \texttt{dyck-2} & Returns a sequence of ones if the sequence is balanced in regards to open and closed parenthesis and curly brackets, otherwise returns zeros. & \texttt{[\{,(,\{,\},\},)]} & \texttt{[1,1,1,1,1,1]} \\
    \bottomrule
    \end{tabular}
    }
    \vspace{1em}
    \caption{
        Base RASP programs from \cite{weiss2021thinking} used to build \raspbugs{}.
        RASP programs handle lists of tokens as input and output, computing different sequence-processing tasks.
    }
    \label{tab:rasp-programs}
\end{table}

Buggy program variants were systematically generated by applying a suite of 15 mutation operators to the correct base RASP programs.
These operators include both generic mutations, which alter common programming constructs (e.g., replacing binary operators, modifying numerical constants), and RASP-specific mutations, tailored to the unique features of the RASP language (e.g., negating selector outputs, incrementing/decrementing RASP indices).
Mutations are applied individually or in combination to create a rich set of buggy programs with varying levels of semantic deviation from the original correct programs.
Details of these mutation operators, including descriptions, examples, the number of occurrences in the generated bugs, and their mean and median accuracy impact, are provided in Table~\ref{tab:mutation-details}.

\begin{sidewaystable}
    \centering
    \resizebox{0.85\textheight}{!}{%
    \begin{tabularx}{\linewidth}{l l X X r r r}
        \toprule
        \textbf{Type} & \textbf{Name} & \textbf{Description} & \textbf{Example} & \textbf{\# Occurrences} & \textbf{Mean Acc.} & \textbf{Median Acc.} \\
        \midrule
        Generic & \texttt{replace-binary-operator} & Replaces a generic binary operator (e.g., "+" to "-") in an expression. & \texttt{x + ...} $\rightarrow$ \texttt{x - ...} & 1055 & 0.17 & 0.00 \\
        Custom & \texttt{replace-rasp-comparison} & Changes a RASP comparison operator (e.g., \texttt{EQ} to \texttt{LT}) in a select statement. & \texttt{Comparison.EQ} $\rightarrow$ \texttt{Comparison.LT} & 896 & 0.26 & 0.00 \\
        Generic & \texttt{replace-comparison-operator} & Replaces a generic comparison operator (e.g., "<" to "<=") in a condition. & \texttt{x < 0} $\rightarrow$ \texttt{x <= 0} & 330 & 0.84 & 0.96 \\
        Custom & \texttt{negate-rasp-sop-select} & Negates a RASP Select operator (e.g., multiplies by -1). & \texttt{rasp.tokens} $\rightarrow$ \texttt{rasp.tokens * -1} & 288 & 0.19 & 0.00 \\
        Generic & \texttt{number-replacer} & Replaces a numeric constant with another value. & \texttt{-1 * ...} $\rightarrow$ \texttt{-0 * ...} & 283 & 0.64 & 0.90 \\
        Custom & \texttt{negate-rasp-sop-constructor} & Negates the result of a RASP SOp constructor. & \texttt{SelectorWidth(...)} $\rightarrow$ \texttt{SelectorWidth(...) * -1} & 160 & 0.24 & 0.00 \\
        Custom & \texttt{decrement-integer} & Decrements an integer constant. & \texttt{min\_key=1} $\rightarrow$ \texttt{min\_key=0} & 140 & 0.57 & 0.82 \\
        Custom & \texttt{increment-integer} & Increments an integer constant. & \texttt{min\_key=1} $\rightarrow$ \texttt{min\_key=2} & 131 & 0.69 & 0.96 \\
        Custom & \texttt{decrement-rasp-indices} & Decrements a RASP indices expression in a select statement. & \texttt{rasp.indices} $\rightarrow$ \texttt{rasp.indices - 1} & 130 & 0.22 & 0.00 \\
        Custom & \texttt{increment-rasp-indices} & Increments a RASP indices expression. & \texttt{rasp.indices} $\rightarrow$ \texttt{rasp.indices + 1} & 127 & 0.25 & 0.00 \\
        Custom & \texttt{negate-rasp-sop-return-stmt} & Negates the return value of a RASP SOp in the return statement. & \texttt{return ...} $\rightarrow$ \texttt{return ... * -1} & 108 & 0.33 & 0.00 \\
        Generic & \texttt{replace-unary-operator} & Changes a unary operator (e.g., "-" to "+") in an expression. & \texttt{-1 * ...} $\rightarrow$ \texttt{+1 * ...} & 74 & 0.45 & 0.04 \\
        Generic & \texttt{zero-iteration-for-loop} & Replaces a for-loop's range with an empty list, skipping the loop. & \texttt{for x in xs:} $\rightarrow$ \texttt{for x in []:} & 19 & 0.93 & 0.96 \\
        Custom & \texttt{negate-rasp-sop-aggregate-value} & Negates the value argumernt of a RASP Aggregate SOp. & \texttt{Aggregate(\dots, sop)} $\rightarrow$ \texttt{Aggregate(\dots, sop * -1)} & 14 & 0.67 & 0.96 \\
        Generic & \texttt{add-not} & Adds a \texttt{not} operator to a condition. & \texttt{x < 0} $\rightarrow$ \texttt{not (x < 0)} & 4 & 0.96 & 0.96 \\
        \bottomrule
    \end{tabularx}%
    }
    \caption{Summary and distribution of mutation operators used in \raspbugs{}. This table details generic and RASP-specific mutation operators, including their descriptions, examples, the number of occurrences in the generated bugs, and their mean and median accuracy impact.}
    \label{tab:mutation-details}
\end{sidewaystable}

\newpage

\section{Repairing Higher-Order Mutants}

\begin{figure}[h]
    \centering
    \includegraphics[width=\linewidth]{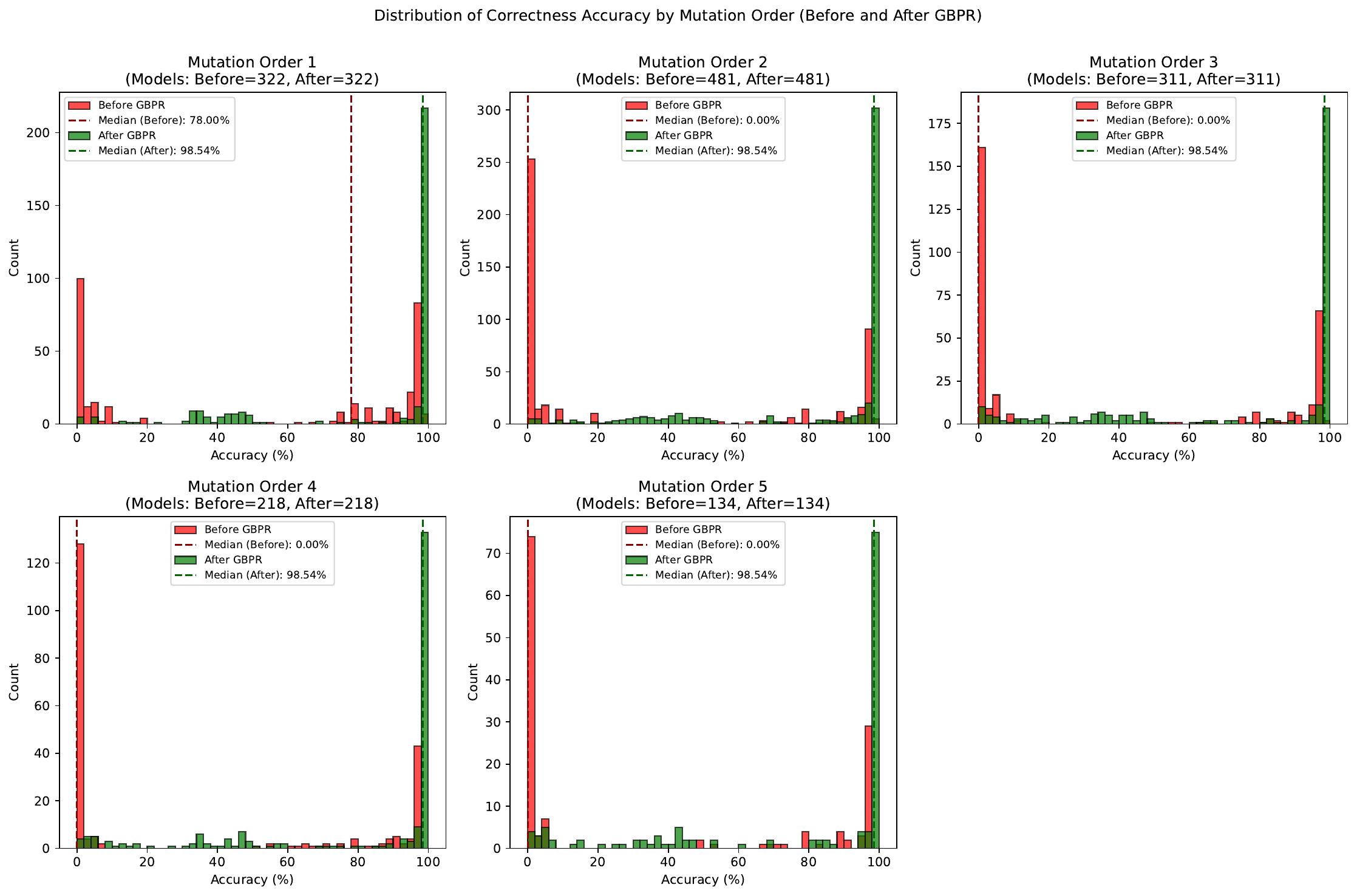}
    \caption{Accuracy distribution before (red) and after (green) \gbprlong{} per mutation order in \raspbugs{}. \gbprlong{} is effective even for higher-order mutants, as shown by the post-repair distributions clustering near 100\% accuracy.}
    \label{fig:accuracy-histogram-by-mutation-order}
\end{figure}

Fixing bugs increases in difficulty as the number of buggy locations increases.
In \raspbugs{}, we generate higher-order mutants by applying multiple mutation operators to the same program.
We evaluate the effectiveness of \gbprlong{} on such mutants in \raspbugs{} by analyzing repair accuracy as a function of mutation order.
As shown in Figure~\ref{fig:accuracy-histogram-by-mutation-order}, \gbprlong{} consistently repairs both single and higher-order mutants, with post-repair accuracy distributions remaining unimodal and concentrated near 100\%.
This suggests that \gbprlong{} is robust to increasing bug complexity and can successfully fix programs even when multiple faults are present.

\newpage

\section{Specification Size Ablation}
\label{app:spec-size}

This appendix presents the full results of the specification-size ablation discussed in \autoref{sec:spec-size}.
We retrain \gbpr{} across $N \in \{100, 1{,}000, 5{,}000, 10{,}000, 25{,}000, 40{,}000\}$ input-output examples, holding the gradient-step budget fixed at 500k steps.
Because \texttt{shuffle\_dyck} only  has 1,635 distinct training samples, including it at $N \geq 5{,}000$ is not possible; to keep a fixed denominator across all $N$ values we exclude it entirely from this ablation, giving 1,135 buggy programs in each condition.
GP and BFS baselines are restricted to the same 1,135 jobs, each with a budget of 500 evaluated programs (matching the original ablation setup; the main experiments in \autoref{sec:experiments} use an extended budget of 1,000).
The validation set is scaled proportionally to $N$: we use $\lfloor N/8 \rfloor$ validation samples (e.g., 5,000 at $N=40{,}000$ and 12 at $N=100$).
The test set is always the fixed held-out 5,000 samples from the original \raspbugs{} specification (as described in \autoref{sec:raspbugs}), ensuring evaluation is comparable across all $N$ values.

\begin{table}[h]
    \centering
    \resizebox{\linewidth}{!}{%
    \begin{tabular}{lrrrrrrr r}
        \toprule
        \textbf{Mutation order} & \textbf{N=100} & \textbf{N=1,000} & \textbf{N=5,000} & \textbf{N=10,000} & \textbf{N=25,000} & \textbf{N=40,000} & \textbf{GP} & \textbf{BFS} \\
        \midrule
        1 (simplest) & 41.2 & 43.3 & 49.4 & 49.4 & 55.9 & 58.0 & 86.9 & \underline{100.0} \\
        2            & 20.3 & 27.1 & 36.3 & 40.8 & 49.7 & 53.7 & 65.3 & \underline{65.8} \\
        3            & 20.5 & 23.9 & 26.9 & 33.3 & 38.0 & 47.0 & \underline{49.1} & 42.3 \\
        4            & 21.6 & 24.4 & 33.0 & \textbf{36.9} & \textbf{49.4} & \textbf{\underline{55.1}} & 35.8 & 36.9 \\
        5 (hardest)  & 23.0 & 24.0 & 25.0 & 34.0 & \textbf{38.0} & \textbf{\underline{43.0}} & 36.0 & 31.0 \\
        \midrule
        All & 25.3 & 29.3 & 35.7 & 39.9 & 47.6 & 52.5 & 59.5 & \underline{60.8} \\
        \bottomrule
    \end{tabular}%
    }
    \caption{
        Percentage of bugs fixed (test accuracy $\geq 99\%$) by specification size $N$ and mutation order.
        \textbf{Bold} entries indicate \gbpr{} values where \gbpr{} surpasses \emph{both} GP and BFS; \underline{underline} marks the row maximum.
        \gbpr{} crosses both symbolic baselines on order-4 bugs at $N=10{,}000$ and on order-5 bugs at $N=25{,}000$.
        For simple bugs (orders 1--2), symbolic methods retain their advantage throughout.
    }
    \label{tab:spec-size}
\end{table}

\begin{figure}[h]
    \centering
    \includegraphics[width=0.48\linewidth]{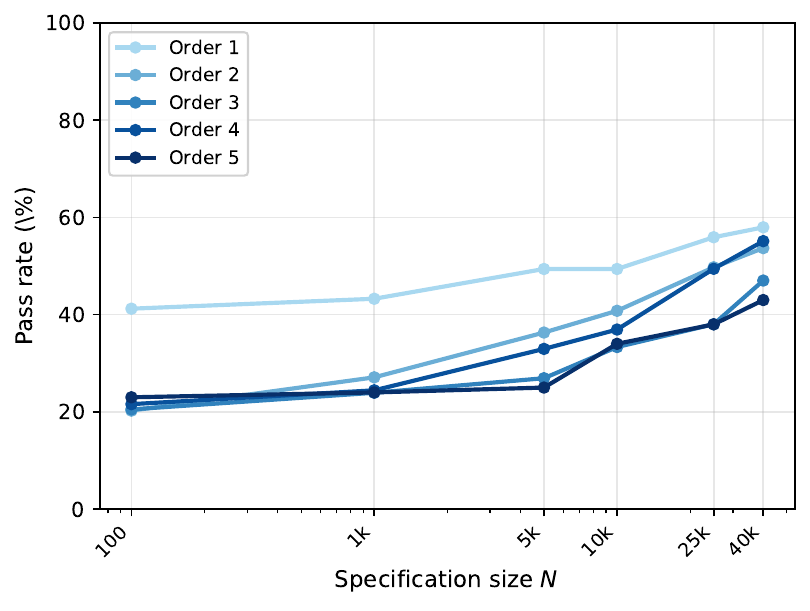}
    \hfill
    \includegraphics[width=0.48\linewidth]{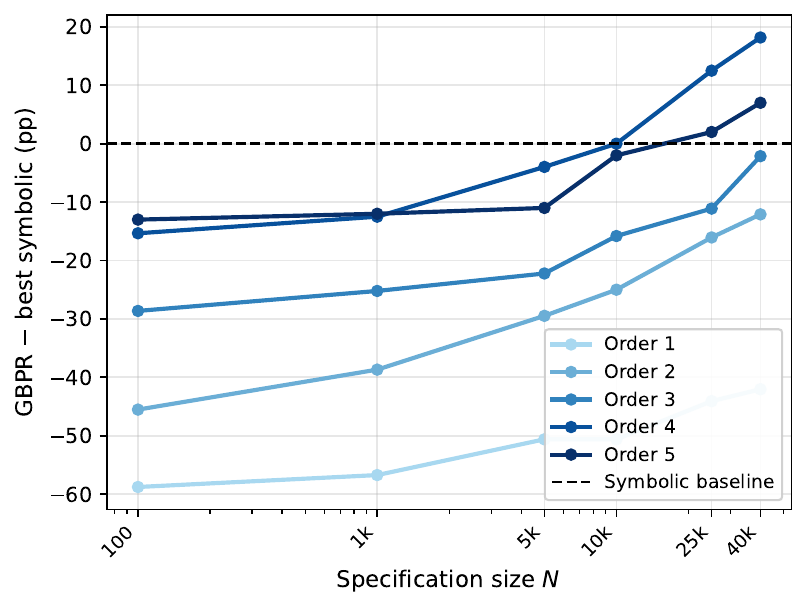}
    \caption{
        \textbf{Left:} Pass rate ($\geq 99\%$ accuracy) of \gbpr{} versus specification size $N$ (log scale), one curve per mutation order.
        \textbf{Right:} \gbpr{} advantage over the best symbolic baseline (\(\max(\text{GP},\text{BFS})\)) per order, in percentage points.
        The dashed line at $y=0$ marks parity with the best symbolic method.
        Orders 4 and 5 cross into positive territory at $N=10{,}000$ and $N=25{,}000$ respectively; orders 1--3 remain below parity throughout the tested range.
    }
    \label{fig:spec-size}
\end{figure}